\documentclass[12pt]{article}

\usepackage{amsmath,amsthm,amssymb,amscd}
\usepackage{graphicx, float}
\usepackage[hidelinks]{hyperref}
\usepackage{natbib}
\usepackage{ds}

\usepackage{tabularx} % automatic line breaks in tables
\newcolumntype{L}[1]{>{\raggedright\arraybackslash}p{#1}} % left fixed width

\usepackage{enumitem} % using indent items in enumerate
\usepackage{caption, subcaption} % using the caption for tables and figures
\usepackage{capt-of} % cross-referencing the figures in appendix
\usepackage[flushleft]{threeparttable} % adding the footnote in a table
\usepackage{multirow}
\usepackage{booktabs}

\usepackage{xcolor,soul} % adding color fonts and highlight text
\usepackage{eurosym}

\usepackage{authblk}

\newtheorem{theorem}{Theorem} %numbering by default: (1), (2),...
\title{\bf Modeling economies of scope in joint production: Convex regression of input distance function}

\author{Timo Kuosmanen, \hspace*{1mm}  Sheng Dai\footnote{Corresponding author. \newline
\hspace*{5mm} \textit{E-mail addresses:}  \texttt{timo.kuosmanen@utu.fi (T. Kuosmanen)}, \texttt{sheng.dai@utu.fi (S. Dai)}.}}

\affil{Department of Economics, Turku School of Economics, University of Turku}

\date{\today}

\begin{document}
% revising the caption Figure XX: to Fig. XX.; Table XX: to Table XX.
\captionsetup[figure]{labelfont={bf},labelformat={default},labelsep=period,name={Fig.}}
\captionsetup[table]{labelfont={bf},labelformat={default},labelsep=period,name={Table}}

\maketitle
 
\vfill
% \begin{center}
% Preprint submitted to \emph{European Journal of Operational Research}
% \end{center}
\vfill

\begin{abstract}
\noindent Modeling of joint production has proved a vexing problem. This paper develops a radial convex nonparametric least squares (CNLS) approach to estimate the input distance function with multiple outputs. We document the correct input distance function transformation and prove that the necessary orthogonality conditions can be satisfied in radial CNLS. A Monte Carlo study is performed to compare the finite sample performance of radial CNLS and other deterministic and stochastic frontier approaches in terms of the input distance function estimation. We apply our novel approach to the Finnish electricity distribution network regulation and empirically confirm that the input isoquants become more curved. In addition, we introduce the weight restriction to radial CNLS to mitigate the potential overfitting and increase the out-of-sample performance in energy regulation.
\\[5mm]
\end{abstract}
\vfill

\thispagestyle{empty}

\newpage
\setcounter{page}{1}
\setcounter{footnote}{0}
\pagenumbering{arabic}
\baselineskip 20pt

%----------------%
%
%----------------%

\section{Introduction}\label{sec:intro}
 
While the conventional microeconomic theory takes a single-product firm as a norm, there is growing empirical evidence that multiproduct firms are more common than usually assumed. \citet{Bernard2010} were among the first to argue that extensive reallocation of resources occurs through product switching by multiproduct firms: product switching is frequent, widespread, and influential in determining both aggregate and firm outcomes. A recent empirical study by \citet{Kuosmanen2023} suggest that approximately one-half of the Finnish manufacturing firms are multi-product firms, and approximately 20 percent of firms operate in multiple industry divisions.

Modeling the joint production of multiple outputs has proved a more vexing problem than most authors realize. In the deterministic setting where the input-output data contain no noise, the technology distance functions such as the input and output distance functions by \citet{Shephard1970} and the directional distance functions (DDF) by \citeauthor{Chambers1996} (\citeyear{Chambers1996}, \citeyear{Chambers1998b}) provide theoretically sound representations of the technology, which can be conveniently estimated by linear programming techniques known as data envelopment analysis (DEA; \citeauthor{Charnes1978}, \citeyear{Charnes1978}).\footnote{
    The distance functions are reciprocals of Farrell's (\citeyear{Farrell1957}) radial measures of technical efficiency. However, the distance functions are representations of technology, which can be used for analyzing marginal properties such as elasticities of substitution, transformation, or scale, even if one is not primarily interested in efficiency measurement. 
}
By contrast, the situation becomes more difficult in the stochastic case, where data are perturbed by random noise. Most approaches known in the literature are based on hidden assumptions that are highly restrictive and unlikely to hold.

In mainstream economics, a commonly used approach to modeling multiproduct firms is to estimate a system of single-output production functions, dividing the firm's total inputs by outputs based on the revenue shares of outputs (\citeauthor{Foster2008}, \citeyear{Foster2008}). However, such a proportional assignment of ratios of inputs to outputs seems completely arbitrary: a profit-maximizing multiproduct firm would demand inputs proportionate to the revenue shares of outputs only under very restrictive assumptions, which are unlikely to hold in the real world. \citet{De2016} propose an improved version of this approach, where they use the data of single-product firms to estimate production functions and then interpolate the estimated functions to model multiproduct firms. However, the linear interpolation approach assumes away possible synergies between multiple products: the resulting output isoquants are linear by construction. This seems counterintuitive due to economies of scope, in which physical synergies imply strictly convex output sets. Indeed, synergies are often seen as the main reason why firms find it beneficial to produce multiple products jointly rather than specialize in a single product (e.g., \citeauthor{Panzar1981}, \citeyear{Panzar1981}). Therefore, assuming away the possible physical synergies from the outset seems restrictive.

In the econometric literature of stochastic frontier analysis (SFA), the modeling of joint production is usually based on Shephard's (\citeyear{Shephard1970}) input and output distance functions (\citeauthor{Lovell1994}, \citeyear{Lovell1994}; \citeauthor{Coelli1999}, \citeyear{Coelli1999}; \citeauthor{Kumbhakar2013}, \citeyear{Kumbhakar2013}). While the distance functions provide a general and theoretically sound framework for modeling joint production, the parametric functional forms usually employed in SFA tend to violate the theoretical properties of the distance function. For instance, it is easy to show that the Cobb-Douglas functional form for the distance function cannot support convex output sets at any parameter values, in fact, the output sets are not even bounded. The commonly used translog functional form is flexible enough to support convex output sets, but in that case, the output sets bend backward, violating the free disposability property. As a result, the translog distance functions cannot satisfy both convexity and free disposability axioms. 

To our knowledge, the only theoretically sound approach to estimating technology distance functions of multiproduct firms is to resort to the shape-constrained nonparametric regression, following \citet{Kuosmanen2017a}. Convex regression is a fully nonparametric method imposing shape constraints such as monotonicity, convexity, and homogeneity. The desired axiomatic properties of distance functions are thus guaranteed to hold. To disentangle technical inefficiency from random noise, \citet{Kuosmanen2017a} employ the nonparametric kernel deconvolution method suggested by \citet{Hall2002}. If the parametric distributional assumptions on inefficiency and/or noise are imposed, one can combine convex regression with the SFA techniques, as first outlined by \citet{Kuosmanen2012c}, who refer to the blended application of nonparametric and parametric techniques as stochastic nonparametric envelopment of data (StoNED).

Thus far, the convex regression and StoNED estimation of distance functions have focused on either DDF by \citet{Kuosmanen2017a} or a prior transformation on output variables by \citet{Schaefer2018}. DDF is a general and theoretically valid representation of the technology, but its additive structure implies a somewhat limiting additive structure for the composite error term. In econometrics, the commonly used log-linear functional forms imply a multiplicative error term, where the size of the error is proportionate to the output. While a prior transformation on output variables is a meaningful attempt, the necessary orthogonality condition in \citet{Schaefer2018} can not be satisfied, and the approach is not theoretically sound, as we demonstrate below.

The purpose of this paper is to present a theoretically sound approach to model economies of scope in joint production using the convex regression approach. Our first methodological contribution is to develop a radial convex nonparametric least squares (CNLS) formulation to estimate the input distance function with the multiple input-multiple output specification. The orthogonality condition in radial CNLS will be stated and proven. We also extend the proposed radial CNLS approach to estimate output distance functions and the expected inefficiency. 

Our second contribution is to compare the finite sample performance of radial CNLS versus the standard DEA and SFA estimators in terms of the input distance function estimation with multiple outputs in the controlled environment of the Monte Carlo study. The simulations reveal the superior performance of the radial CNLS estimator in the joint production setting involving multiple outputs.

The third contribution is to empirically apply the radial CNLS estimator to the Finnish electricity distribution firms' dataset. In contrast to the earlier DDF approach (\citeauthor{Kuosmanen2017a}, \citeyear{Kuosmanen2017a}) or the CNLS with input requirement function approach (\citeauthor{Kuosmanen2020d}, \citeyear{Kuosmanen2020d}), the proposed radial CNLS approach does not rely on any prior specification of the direction vector. To reduce the potential overfitting in shadow prices and increase the out-of-sample performance in energy regulation, we introduce the weight restriction constraints to radial CNLS. The results confirm that the input isoquants become more curved, implying better substitution possibilities. if we use two inputs rather than just project in the direction of one input as in the input requirement function. These results are highly relevant for the incentive regulation of electricity distribution firms in the real world.

The rest of this paper is organized as follows. Section \ref{sec:theory} introduces the theory on the input distance function and the regression model. Section \ref{sec:convex} presents a naive approach to estimating the input distance function and proposes the radial CNLS approach. The possible extensions are discussed in Section \ref{sec:exten}. The Monte Carlo study and the empirical application of Finland's electricity distribution network regulation are demonstrated in Sections \ref{sec:mc} and \ref{sec:app}. Section \ref{sec:concl} concludes this paper with suggested avenues for future research. The formal proofs of Theorems and additional tables and figures are provided in the Online Supplement.

%-----------------%
% 
%-----------------%

\section{Theory}\label{sec:theory}

\subsection{Input distance function}

Consider a sample of firms where at least some are multiproduct firms, but a subset may specialize in just one or few products. We assume that each firm takes the output demands as given, and there is no noise in the observed outputs $\by \in \real_+^{s}$. The firm managers optimize some meaningful objective functions, e.g., minimizing cost. Note that cost minimization is equivalent to profit maximization when the output demands are taken as given. The observed inputs $\bx \in \real_+^{m}$ are subject to random perturbances $\varepsilon_i$, and the firm manager then adjusts their input demands according to 
\begin{equation}
	\bx_i = \exp(\varepsilon_i)\bx^*
    \label{eq:dgp}
\end{equation}
where $\bx^*$ is the optimal (cost minimizing) input vector. 

Consider further the following production possibility set with the generated data $(\bx, \by)$.
\begin{equation}
	T = \{(\bx, \by) \in \real_{+}^{m+s} \mid \bx \text{ can produce } \by \}
\end{equation}
where we assume that $T$ is a closed and nonempty set satisfying free disposability (monotonicity) and convexity axioms. See \citet{Fare1995} for further detailed discussions on the axiomatic production theory. 

The Shephard's input distance function $D_i: \real_+^{m} \times  \real_+^{s} \rightarrow \real_+^{1} \cup \{+ \infty\}$ is defined as \citep{Shephard1970}:
\begin{equation}
	D_i(\bx, \by) = \sup\{\phi \mid (\bx/\phi, \by) \in T\}
    \label{eq:df}
\end{equation}
where the construction of input distance function $D_i$ does not depend on behavioral hypotheses such as profit maximization or cost minimization.

Similar to \citet{Shephard1970}, we consider $D_i$ as a general functional representation of multi-output technology: If $D_i(\bx, \by)=1$, then production plan $(\bx, \by)$ is technically feasible and efficient. If $D_i(\bx, \by)<1$, then $(\bx, \by)$ is technically feasible but inefficient. If $D_i(\bx, \by)>1$, then $(\bx, \by)$ is technically infeasible. Therefore, $(\bx, \by)$ for which $D_i(\bx,\by) \le 1$ are within the production possibility set and $(\bx, \by)$ for which $D_i(\bx, \by)=1$ are its efficient subset. 

Linear homogeneity of $D_i$ is a fundamental property that does not relate to the properties of $T$, but arises from the definition of $D_i$. The linear homogeneity implies that 
\begin{equation}
	D_i(\lambda \bx, \by) = \lambda D_i(\bx, \by)\quad \forall \lambda > 0, (\bx, \by) \in T
    \label{eq:home}
\end{equation}
This property helps develop the following convex regression model of the input distance function. The value of the input distance function $D_i(\bx_i, \by_i)$ is then completely determined by the error term $\varepsilon_i$ (see Theorem \ref{the:the1}).
\begin{theorem}
If the observed data are generated according to the data generating process (DGP) described, then the value of the input distance function is equal to
	\begin{equation*}
		D_i(\bx_i, \by_i) = \exp(-\varepsilon_i)
	\end{equation*}
	\label{the:the1}
\end{theorem}
\vspace{-1.6cm}
\begin{proof}
	See Appendix A.
\end{proof} 

This result is analogous to Proposition 2 in \citet{Kuosmanen2017a}, adapted to the present setting of $D_i$. Note that even though the distance function is a deterministic representation of a deterministic technology, its value is a random variable when the observed data are perturbed by random noise and inefficiency.

\subsection{Regression model}

To derive the regression equation, we can utilize the homogeneity property \eqref{eq:home} and normalize by $x_{1i}$ to obtain\footnote{
    Note that any one of the input variables can be selected in the distance function normalization.
}
\begin{equation}
	D_i(\bx_i/x_{1i}, \by_i) = (1/x_{1i}) \cdot D_i(\bx_i, \by_i)
    \label{eq:regeq1}
\end{equation}
where $\bx/x_1 = (1, x_2/x_1, \ldots, x_m/x_1)^\prime$. After taking the logs, we have $\ln x_{1i} = -\ln D_i(\bx_i/x_{1i}, \by_i) + \ln D_i(\bx_i, \by_i)$. 

Using Theorem \ref{the:the1}, we substitute $\ln D_i(\bx_i, \by_i)$ with the error term $\varepsilon_i$ and obtain
\begin{equation}
	\ln x_{1i} = -\ln D_i(\bx_i/x_{1i}, \by_i) + \varepsilon_i
 \label{eq:regeq2}
\end{equation}

A similar regression equation is commonly used in the SFA literature (e.g., \citeauthor{Kumbhakar2013}, \citeyear{Kumbhakar2013}). The main difference is that we derive the regression equation without imposing any particular functional form for $D_i$. Note that the parametric functional forms usually employed in SFA tend to violate the theoretical properties of the distance function such as monotonicity and convexity of the output sets. In the SFA literature, $\varepsilon_i$ consists of asymmetric inefficiency $u \sim N^+(0, \sigma_u^2)$ and symmetric noise $v \sim N(0, \sigma_v^2)$. For simplicity, we here mainly focus on the conventional case of a symmetric error term, but the extension to stochastic frontier setting with an asymmetric composite error term is briefly examined in section~\ref{sec:eff} below. 

%-----------------%
% 
%-----------------%
\vspace{-0.43cm}
\section{Convex regression}\label{sec:convex}
\subsection{A naive approach}

To estimate the regression equation \eqref{eq:regeq2} developed in the previous section by convex regression, it would be tempting to formulate the CNLS problem as follows
\begin{alignat}{2}
	\underset{\chi, \alpha, \bbeta, \bgamma, \varepsilon} {\mathop{\min }}\, \quad & \sum\limits_{i=1}^{N}\varepsilon_i^2 &{\quad}& \label{eq:cnls_naive}\\
	\mbox{s.t.}\quad
    & \ln x_{1i} = - \ln (\chi_i)  + \varepsilon_{i}  &{}& \forall i\notag \\
	& \chi_i = \alpha_i + \bbeta^{\prime}_i (\bx_i/x_{1i}) - \bgamma^{\prime}_i \by_i &{}& \forall i\notag \\
	& \chi_i \le \alpha_h + \bbeta^{\prime}_{h}  (\bx_i/x_{1i}) - \bgamma^{\prime}_{h} \by_i &{}& \forall i,h \notag  \\ 
	& \bbeta_i \ge \textbf{0}, \bgamma_i \ge \textbf{0} &{}& \forall i\notag  
\end{alignat}

The first two constraints of \eqref{eq:cnls_naive} are used to characterize the regression equation \eqref{eq:regeq2}. The decision variable $\chi_i$ denotes the distance function $D_i(\bx_i/x_{1i}, \by_i)$. The second constraint can thus be interpreted as a multivariate linear regression among the normalized inputs, outputs, and distance function. The third constraint imposes the concavity on the distance function. The last set of constraints guarantees the monotonicity of the distance function. 

Problem \eqref{eq:cnls_naive} is henceforth referred to as a naive CNLS because the estimated CNLS residuals do not satisfy the usual sample orthogonality condition of the linear regression. In the parametric regression, the linear homogeneity of $D_i$ is satisfied by construction because the regression residuals $\hat{\varepsilon}$ satisfy the following sample orthogonality condition
\begin{equation*}
    \sum{(\bx_i/x_{1i}})\cdot \hat{\varepsilon}_i=0
\end{equation*}
\vspace{-0.1cm}
which is the sample counterpart of the population orthogonality condition
\begin{equation*}
\E[(\bx_i/x_{1i})\cdot\varepsilon_i]=0.
\end{equation*}

This implies the equiproportionate radial orientation of $D_i$, but as noted above, the CNLS residuals do not generally satisfy these orthogonality conditions by construction. Therefore, we need to find a way of imposing the orthogonality for the input ratios $(\bx_i/x_{1i})$.

We note that \citet{Schaefer2018} propose an extended CNLS approach to estimate the ray production function in the multiple input-multiple output specification. This approach relies on the polar coordinate transformation to introduce the angle-related variables and generates a ``multiple input-single output'' model. While it is a meaningful attempt to model multiple outputs in the framework of CNLS, similar to problem \eqref{eq:cnls_naive}, the proposed formulation by \citet{Schaefer2018} does not impose the orthogonality conditions and therefore fails to satisfy the linear homogeneity property of the radial distance functions. 

\subsection{Proposed approach}

To gain intuition, consider first the special case of two inputs. We utilize the homogeneity property and normalize inputs by the Cobb-Douglas function $(x_{1i}^{1-d}x_{2i}^d)$, $0 < d < 1$, to obtain
\begin{equation}
    D_i(\bx_i/(x_{1i}^{1-d}x_{2i}^d), \by_i)=(x_{1i}^{d-1}x_{2i}^{-d})\cdot D_i(\bx_i,\by_i)
    \label{eq:regeq3}
\end{equation}
where we can use any arbitrary parameter value $d$ (e.g., $d = 0.5$). We will later ensure that the specific choice of $d$ does not matter. 

Taking logs and reorganizing, we now have the following regression equation
\begin{equation}
    \ln x_{1i} = -\ln D_i(\bx_i/(x_{1i}^{1-d}x_{2i}^d), \by_i) + d(\ln x_{1i} - \ln x_{2i}) + \varepsilon_i
    \label{eq:regeq4}
\end{equation}

To estimate the equation \eqref{eq:regeq4}, we propose the following radial CNLS approach to model multiple input-multiple output setting
\begin{alignat}{2}
	\underset{\chi, \alpha, \bbeta, \bgamma, \delta, \varepsilon} {\mathop{\min }}\, \quad & \sum\limits_{i=1}^{N}\varepsilon_i^2 &{\quad}& \label{eq:cnls_raidal}\\
	\mbox{s.t.}\quad
    & \ln x_{1i} = - \ln (\chi_i) + \delta(\ln x_{1i} - \ln x_{2i}) + \varepsilon_{i}  &{}& \forall i\notag \\
	& \chi_i = \alpha_i + \bbeta^{\prime}_i (\bx_i/(x_{1i}^{1-d}x_{2i}^d)) - \bgamma^{\prime}_i \by_i &{}& \forall i\notag \\
	& \chi_i \le \alpha_h + \bbeta^{\prime}_{h}  (\bx_i/(x_{1i}^{1-d}x_{2i}^d)) - \bgamma^{\prime}_{h} \by_i &{}& \forall i,h \notag  \\ 
	& \bbeta_i \ge \textbf{0}, \bgamma_i \ge \textbf{0} &{}& \forall i\notag  
\end{alignat}
where problem \eqref{eq:cnls_raidal} is a semi-nonparametric partial linear formulation building on \citeauthor{Johnson2011} (\citeyear{Johnson2011}, \citeyear{Johnson2012a}). Similar to problem \eqref{eq:cnls_naive}, the first two sets of constraints to represent the regression equation \eqref{eq:regeq4}, the last two sets of constraints impose the concavity and monotonicity on the input distance function, respectively. 

In contrast to parameter $d$ that we simply postulate in equation \eqref{eq:regeq3}, $\delta$ is a decision variable to be estimated. Theorem \ref{the:the2} shows that the semi-nonparametric partial linear formulation satisfies the orthogonality condition and hence $\delta$ can be a free decision variable in problem \eqref{eq:cnls_raidal}.
\begin{theorem}
The radial CNLS residuals obtained as the optimal solution to problem \eqref{eq:cnls_raidal} satisfy the following sample orthogonality condition
	\begin{equation*}
		\sum_{i=1}^{n}(\ln x_{1i} - \ln x_{2i})\cdot \hat{\varepsilon}_i=0
	\end{equation*}
	\label{the:the2}
\end{theorem}
\vspace{-1.6cm}
\begin{proof}
	See Appendix A.
\end{proof} 

Theorem \ref{the:the2} shows that problem \eqref{eq:cnls_raidal} effectively enforces the equiproportionate radial orientation, in other words, the linear homogeneity, which is an essential property of $D_i$. We thus can consistently estimate the regression equation \eqref{eq:regeq4} using the nonparametric radial CNLS formulation \eqref{eq:cnls_raidal}. 

Regarding the choice of parameter $d$, it is clearly possible or perhaps even likely that the coefficient $\delta$ obtained as the optimal solution to problem \eqref{eq:cnls_raidal} differs from our arbitrary specification of $d$. But it is unnecessary to guess the optimal value of $d$ beforehand. If we solve the problem \eqref{eq:cnls_raidal} for multiple different values of $d$, it is easy to verify that the parameter estimates of $\delta$ change, however, the residuals $\hat{\varepsilon}_i$ remain exactly the same irrespective of the choice of $d$. It turns out that $\delta$ is an over-identified parameter value that does not influence the optimal solution. More specifically, we need to consider $-\ln(\chi_i)+\delta(\ln x_{1i} - \ln x_{2i})$ as the estimate of $D_i$, not only the nonparametric part $-\ln(\chi_i)$. Note that the sum of two concave functions is itself a concave function. As the value of parameter $\delta$ changes, so does that of the nonparametric part, but their sum remains the same. This can be referred to as partial identification in the present context (see, e.g., \citeauthor{Manski2009}, \citeyear{Manski2009}; \citeauthor{Tamer2010}, \citeyear{Tamer2010}). 

To find a unique estimate of $D_i$ in any given point $(\bx_i, \by_i)$, we can solve the following linear programming (LP) problem, directly analogous to the multiplier formulation of DEA

\begin{alignat}{2}
	 \hat{D}_i&(\bx_i,\by_i) = \max \bgamma^{\prime} \by_i  &{\quad}& \label{eq:lp}\\
	\mbox{s.t.}\quad
    & \bbeta^{\prime} \exp(-\hat{\varepsilon}_i) \bx_i = 1  &{}& \forall i\notag \\
	& \alpha_i+\bbeta^{\prime} \exp(-\hat{\varepsilon}_h)\bx_h-\bgamma^{\prime}\by_h \ge 0  &{}& \forall i, h\notag \\
	& \bgamma \ge \textbf{0}, \bbeta \ge \textbf{0} &{}& \forall i\notag  
\end{alignat}

The adjustment by the CNLS residuals $\hat{\varepsilon}_i$ of problem \eqref{eq:cnls_raidal} effectively projects all observations to the estimated production frontier (in this case the average practice technology). The optimal solution to this LP problem provides the shadow prices $\bgamma$ and $\bbeta$ associated with the minimum extrapolation technology.

%-----------------%
% 
%-----------------%

\section{Extensions}\label{sec:exten}

In this section, we consider the general case with more than two inputs, connect with the efficiency analysis, and estimate the output distance function.

\subsection{More than two inputs}

To generalize the results of the previous section, consider the general case of $M$ inputs (i.e., $M > 2$). Let us normalize $D_i$ by the Cobb-Douglas function  
\begin{equation*}
     (x_{1i}^{1/M}x_{2i}^{1/M}\ldots x_{Mi}^{1/M}) = (x_{1i}^{1-(M-1)/M}x_{2i}^{1/M}\ldots x_{Mi}^{1/M}).
\end{equation*}

In this case, the regression equation \eqref{eq:regeq2} becomes 
\begin{equation}
    \ln x_{1i} = -\ln D_i(\bx_i/(x_{1i}^{\frac{1}{M}}x_{2i}^{\frac{1}{M}}\ldots x_{Mi}^{\frac{1}{M}}), \by_i)+\dfrac{1}{M}\sum_{m=2}^{M} (\ln x_{1i}- \ln x_{mi})+\varepsilon_i
    \label{eq:regeq5}
\end{equation}
and the radial CNLS estimator can be stated as 
\vspace{-0.09cm}
\begin{alignat}{2}
	\underset{\chi, \alpha, \bbeta, \bgamma, \delta, \varepsilon} {\mathop{\min }}\, \quad & \sum\limits_{i=1}^{N}\varepsilon_i^2 &{\quad}&  \label{eq:cnls3}\\
	\mbox{s.t.}\quad
    & \ln x_{1i} = - \ln (\chi_i) + \sum_{m=2}^{M} \delta_m(\ln x_{1i} - \ln x_{mi}) + \varepsilon_{i}  &{}& \forall i\notag \\
	& \chi_i = \alpha_i + \bbeta^{\prime}_i (\bx_i/(x_{1i}^{\frac{1}{M}}x_{2i}^{\frac{1}{M}}\ldots x_{Mi}^{\frac{1}{M}})) - \bgamma^{\prime}_i \by_i &{}& \forall i\notag \\
	& \chi_i \le \alpha_h + \bbeta^{\prime}_{h}  (\bx_i/(x_{1i}^{\frac{1}{M}}x_{2i}^{\frac{1}{M}}\ldots x_{Mi}^{\frac{1}{M}})) - \bgamma^{\prime}_{h} \by_i &{}& \forall i,h \notag  \\ 
	& \bbeta_i \ge \textbf{0}, \bgamma_i \ge \textbf{0} &{}& \forall i\notag  
\end{alignat}

This formulation is a more general CNLS estimation of the input distance function with multiple outputs. That is, problem \eqref{eq:cnls_raidal} is a special case of problem \eqref{eq:cnls3}, where $M=2$. Having solved the radial CNLS problem  \eqref{eq:cnls3}, the same LP problem \eqref{eq:lp} can be applied to obtain the unique estimate of $D_i$. 

\subsection{Efficiency analysis}\label{sec:eff}

Given the residuals $\hat{\varepsilon}_i$ estimated by radial CNLS \eqref{eq:cnls_raidal}, it is possible to estimate the expected inefficiency using the nonparametric kernel deconvolution or if one imposes further parametric distributional assumptions by quasi-likelihood or the method of moments. Following \citet{Kuosmanen2012c} and \citet{Kuosmanen2017a}, we in this section briefly summarize how to estimate the unconditional expected inefficiency $\E(u)$ and the conditional expected inefficiency $\E(u_i \mid \hat{\varepsilon}_i)$.

Nonparametric kernel deconvolution is a full nonparametric estimation of the expected inefficiency $\E(u)=\mu$. For the input distance function, the residuals $\hat{\varepsilon}_i$ are the consistent estimators of $\hat{\varepsilon}_i = \mu + (v_i - u_i)$. The density function of $\hat{\varepsilon}_i$ is defined as 
     \begin{align*}
         \hat{f}_{\hat{\varepsilon}}(x) = (nh)^{-1} \sum_{i=1}^{n}K\bigg(\frac{x-\hat{\varepsilon}_i}{h} \bigg)
     \end{align*}
where $K(\cdot)$ is a standard kernel function, $h$ is the bandwidth, $x$ is the projected data of $\hat{\varepsilon}$, and $n$ is the number of observations. The first derivative of the density function of the composite error term $(f_{\hat{\varepsilon}}^{\prime})$ is proportional to that of the inefficiency term ($f_u^{\prime}$) in the neighborhood of $\mu$ (\citeauthor{Hall2002}, \citeyear{Hall2002}). Therefore, $\hat{\mu} = \arg \max_{x \in C}(\hat{f}_{\hat{\varepsilon}}^{\prime}(x))$ provides a nonparametric estimation of the expected inefficiency $\mu$, where $C$ is a closed interval in the right tail of $f_{\varepsilon}(\cdot)$.

Alternatively, the method of moments and the quasi-likelihood approaches for the residual decomposition build upon additional parametric distributional assumptions (e.g., the half-normal inefficiency and normal noise). For the method of moments approach, we first calculate the second and third central moments of the residuals $\hat{\varepsilon}_i$ by
    \begin{align*}
     \hat{M_2} &= \sum_{i=1}^{n}(\hat{\varepsilon}_i-\bar{\varepsilon})^{2}/n  \\
     \hat{M_3} &= \sum_{i=1}^{n}(\hat{\varepsilon}_i-\bar{\varepsilon})^{3}/n  
    \end{align*}

Under the parametric distributional assumptions, we have $M_3 = (2/\pi)^{1/2}(1-4/\pi)\sigma_u^2$ and $M_2 = \sigma_v^2+\sigma_u^2(\pi-2)/\pi$. Given the estimated $\hat{M}_2$ and $\hat{M}_3$, we then calculate $\sigma_u$ and $\sigma_v$. Formally, 
     \begin{align*}
         \hat{\sigma}_u &= \sqrt[3]{\frac{\hat{M_3}}{\bigg(\sqrt{\frac{2}{\pi}}\bigg)\bigg[1-\frac{4}{\pi}\bigg]}} \\
         \hat{\sigma}_v &= \sqrt[2]{\hat{M_2}-\bigg[\frac{\pi-2}{\pi}\bigg] \hat{\sigma}_u^2 }
     \end{align*}

The quasi-likelihood estimation is an alternative to compute $\hat{\sigma}_u$ and $\hat{\sigma}_v$. We apply the standard maximum likelihood method to estimate the following likelihood function
     \begin{align*}
         \ln L(\lambda) & = -n\ln(\hat{\sigma}) + \sum \ln \Phi\bigg[\frac{-\varepsilon_i \lambda}{\hat{\sigma}}\bigg] - \frac{1}{2\hat{\sigma}^2}\sum\varepsilon_i^2 
     \end{align*}
where
     \begin{align*}
        \varepsilon_i &= \hat{\varepsilon}_i-(\sqrt{2}\lambda\hat{\sigma})/[\pi(1+\lambda^2)]^{1/2}    \\
        \hat{\sigma} &= \Bigg\{\frac{1}{n} \sum(\hat{\varepsilon}_i)^2 / \bigg[1 - \frac{2\lambda^2}{\pi(1+\lambda^2)}\bigg] \Bigg\}  
     \end{align*}

The likelihood function $\ln L(\lambda)$ consists of a single parameter $\lambda$. $\Phi$ denotes the cumulative distribution function of normal distribution. After obtaining $\hat{\sigma}_u$ by these two parametric approaches, we can compute the expected inefficiency, $\mu = (2/\pi)^{1/2}\hat{\sigma}_u$.

\citet{Jondrow1982} propose a widely applied estimator for calculating the conditional expected inefficiency $\E[u_i \mid \hat{\varepsilon}_i]$. Specifically, for the input distance function, the conditional expected value of inefficiency $\E[u_i \mid \varepsilon_i]$ is formulated as
\begin{align*}
         \E[u_i \mid \hat{\varepsilon}_i]
         &= \mu_{*i} + \sigma_* \Bigg[ \frac{\phi(-\mu_{*i}/\sigma_*)}{1-\Phi(-\mu_{*i}/\sigma_*)} \Bigg] = \sigma_* \Bigg[ \frac{\phi(\hat{\varepsilon}_i \lambda/\sigma)}{1-\Phi(\hat{\varepsilon}_i \lambda/\sigma)} - \frac{\hat{\varepsilon}_i \lambda}{\sigma} \Bigg]
\end{align*}
where $\mu_{*i}= -\hat{\varepsilon}_i \sigma_u^2/\sigma^2$, $\sigma_*^2 = \sigma_u^2\sigma_v^2/\sigma^2$, $\lambda = \sigma_u/\sigma_v$, and $\sigma^2 = \sigma_u^2 +\sigma_v^2$. $\phi$ and $\Phi$ are the standard normal density function and its cumulative distribution function, respectively. Note that the conditional mean inefficiency in \citet{Jondrow1982} could be further extended to conditional quantile inefficiency. 
    
\subsection{Output distance function}

Consider now the general output distance function
\begin{equation}
	D_i^O(\bx, \by) = \inf\{\phi \mid (\bx, \by/\phi) \in T\}
    \label{eq:df2}
\end{equation}
where $D_i^O(\bx, \by)$ is homogeneous of degree 1 in $\by$, implying that 
\begin{equation*}
	D_i^O(\bx, \lambda \by) = \lambda D_i^O(\bx, \by)\quad \forall \lambda > 0, (\bx, \by) \in T.
\end{equation*}

Given the case of $S$ outputs (i.e., $S \ge 2$) and the linear homogeneity, similar to equation \eqref{eq:regeq5}, the regression equation can be formulated as
\begin{equation}
    \ln y_{1i} = -\ln D_i^O(\bx_i, \by_i/(y_{1i}^{\frac{1}{S}}y_{2i}^{\frac{1}{S}}\ldots y_{Si}^{\frac{1}{S}}))+\dfrac{1}{S}\sum_{s=2}^{S} (\ln y_{1i}- \ln y_{si})+\varepsilon_i
    \label{eq:regeq6}
\end{equation}

We then estimate the output distance function using the following radial CNLS formulation
\begin{alignat}{2}
	\underset{\chi, \alpha, \bbeta, \bgamma, \delta, \varepsilon} {\mathop{\min }}\, \quad & \sum\limits_{i=1}^{N}\varepsilon_i^2 &{\quad}&  \label{eq:cnls4}\\
	\mbox{s.t.}\quad
    & \ln y_{1i} = -\ln (\chi_i) + \sum_{s=2}^{S} \delta_s(\ln y_{1i} - \ln y_{si}) + \varepsilon_i  &{}& \forall i\notag \\
	& \chi_i= \alpha_i + \bgamma^\prime_i(\by_i/(y_{1i}^{\frac{1}{S}}y_{2i}^{\frac{1}{S}}\ldots y_{Si}^{\frac{1}{S}})) - \bbeta^\prime_i \bx_i   &{}& \forall i\notag \\
	& \chi_i \ge \alpha_h +\bgamma^\prime_h(\by_i/(y_{1i}^{\frac{1}{S}}y_{2i}^{\frac{1}{S}}\ldots y_{Si}^{\frac{1}{S}}))-\bbeta^\prime_h\bx_i &{}& \forall i,h \notag  \\ 
	& \bbeta_i \ge \textbf{0}, \bgamma_i \ge \textbf{0} &{}& \forall i\notag  
\end{alignat}
where the first two constraints characterize the regression model \eqref{eq:regeq6}. In contrast to problem \eqref{eq:cnls_raidal}, the third set of constraints of problem \eqref{eq:cnls4} impose the convexity of the output distance function. The last set of constraints also ensures the monotonicity of the output distance function. 

The main challenge here is that the output isoquants of the nonparametric part $-\ln(\chi_i)$ are concave by construction, whereas those of the parametric part $\sum_{s=2}^{S}\delta_s(\ln y_{1i} - \ln y_{si})$ are convex by default. In contrast to the input distance function where both parts have the same curvature, in this case, there is no guarantee that the estimated output sets satisfy convexity. 

If the data are well-behaved, the nonparametric part is flexible enough to offset the wrong curvature of the Cobb-Douglas part. In other words, the formulation \eqref{eq:cnls4} could still satisfy the convexity of the output sets, however, this cannot be guaranteed always to hold. Note that possible violations of convexity of the output sets can always be fixed in the next step, where we apply the LP problem \eqref{eq:lp} to estimate the output distance function. We leave a more detailed examination of the output distance function as an interesting avenue for future research.

%-----------------%
% 
%-----------------%

\section{Monte Carlo study}\label{sec:mc}

In the Monte Carlo study, our main objective is to investigate the finite sample performance of the proposed radial CNLS approach and compare it with the naive CNLS approach and other commonly seen deterministic and stochastic frontier estimation approaches in estimating the input distance function.

\subsection{Setup}

We generate the two input-two output production datasets by using the following two sets of DGPs (\citeauthor{Fare2010}, \citeyear{Fare2010})
\begin{align*}
\begin{split}
    & \text{DGP I} :  P(\bx)=\{(y_1, y_2): y_2 = \beta_0 + \beta_1y_1 + \beta_2y_1^2 + \beta_3y_1^3 + \beta_4y_1^4 + x_1^{0.9}x_2^{0.8} + \varepsilon_i \} 
\end{split}\\
\begin{split}
    &\text{DGP II} : P(\bx)=\{(y_1, y_2): y_2 = \exp(\beta_0 + \beta_1 \ln(y_1) + \beta_2[\ln(y_1)]^2 + \beta_3[\ln(y_1)]^3 + \\
        & \hspace*{6.7cm} \beta_4[\ln(y_1)]^4 + x_1^{0.9}x_2^{0.8}) + \varepsilon_i \} 
\end{split}
\end{align*}
where the inputs $\bx$ in each DGP are independently and randomly drawn from the uniform distribution, $U[0, 1]$, and the error term $\varepsilon$ has three different specifications: $\varepsilon = v$ (i.e., only noise), $\varepsilon = -u$ (i.e., only inefficiency), and $\varepsilon = v - u$ (i.e., both noise and inefficiency). We generate noise $v \sim N(0, \sigma_v^2)$ and inefficiency $u \sim N^+(0, \sigma_u^2)$.

The first output $y_1$ in DGP I is randomly generated from the following two different gamma distributions
\begin{alignat*}{2}
	&\text{Type-A}: y_1 \sim \Gamma(\alpha=5, \beta=0.5); \\
	&\text{Type-B}: y_1 \sim \Gamma(\alpha=18,\beta=0.25)
\end{alignat*}
where $\alpha$ and $\beta$ are the shape and rate parameters of the gamma distribution. For DGP II, the first output $y_1$ is drawn from $U[e^{0.7}, e^{1.4}]$. 

Given the different true parameter combinations ($\beta_k$, $k=0,\ldots,4$) in Table~\ref{tab:tab1}, DGPs I and II consist of three polynomial and translog technologies, respectively. We thus consider 9 models with DGPs I and II in the experiments. Note that Model 3 in each DGP makes the true functions in terms of outputs more concave than the other two models, and Type-B in DGP I generates a more balanced outputs dataset (i.e., the value of generated $y_1$ is close to that of $y_2$).
\begin{table}[H]
	\centering
	\caption{The true parameters used in DGPs I and II.}
		\begin{tabular}{lrrrrrrr}
			\toprule
			& \multicolumn{3}{c}{DGP I} & & \multicolumn{3}{c}{DGP II} \\
			\cmidrule{2-4}\cmidrule{6-8}  & Model 1 & Model 2 & Model 3 && \multicolumn{1}{l}{Model 1} & \multicolumn{1}{l}{Model 2} & \multicolumn{1}{l}{Model 3} \\
			\midrule
			$\beta_0$ & \multicolumn{1}{r}{10.70} & \multicolumn{1}{r}{10.10} & \multicolumn{1}{r}{9.60} && 3.000 & 2.845 & 2.690 \\
			$\beta_1$ & \multicolumn{1}{r}{-0.91} & \multicolumn{1}{r}{-0.72} & \multicolumn{1}{r}{-0.54} && -3.500 & -3.400 & -3.300 \\
			$\beta_2$ &  0.50 $\times$ 10$^{-5}$ &  0.50 $\times$ 10$^{-4}$ & 0.10 $\times$ 10$^{-2}$ && 3.900 & 4.000 & 4.100 \\
			$\beta_3$ & 0.10 $\times$ 10$^{-4}$ & 0.10 $\times$ 10$^{-3}$ &  0.10 $\times$ 10$^{-2}$ && -1.500 & -1.475 & -1.415 \\
			$\beta_4$ & -0.45 $\times$ 10$^{-3}$ & -0.12 $\times$ 10$^{-2}$ & -0.24 $\times$ 10$^{-2}$ && -0.140 & -0.220 & -0.330 \\
			\bottomrule
            \\[-1.5em]
             \multicolumn{8}{l}{\footnotesize \textit{Source}: \citet{Fare2010}.} 
		\end{tabular}%
	\label{tab:tab1}%
\end{table}%
\vspace{-0.5cm}

In all experiments that follow, we resort to the Python/pyStoNED package (\citeauthor{Dai2021b}, \citeyear{Dai2021b}) to solve the CNLS and DEA models and the Python/pySFA package to estimate the SFA models.\footnote{
    A Python Package for Stochastic Frontier Analysis (pySFA): \url{https://github.com/gEAPA/pySFA}.
}
Note that the linear and nonlinear programming problems are solved by the off-the-shelf solvers MOSEK (9.3.11) and KNITRO (13.2), respectively. All experiments are run on Finland's high-performance computing cluster Puhti with Xeon @2.1 GHz processors, 2 CPUs, and 5 GB of RAM per task.

\subsection{Experiment 1}

In experiment 1 we compare the performance of the proposed radial CNLS formulations \eqref{eq:cnls_raidal} and \eqref{eq:cnls3} with the naive CNLS formulation \eqref{eq:cnls_naive} in the absence of inefficiency (i.e., $\varepsilon = v$). We thus consider 135 scenarios with different numbers of observations, $n \in \{25, 50, 100, 200, 400\}$, and the variation of noise levels, $\sigma_v \in \{0.075, 0.15, 0.3\}$. To evaluate their finite sample performance, each scenario is run 500 times to calculate mean square error (MSE) and mean absolute deviation (MAD) (see the Online Supplement) in terms of input distance function estimation.

\begin{figure}[H]
    \centering
    \includegraphics[width=\textwidth]{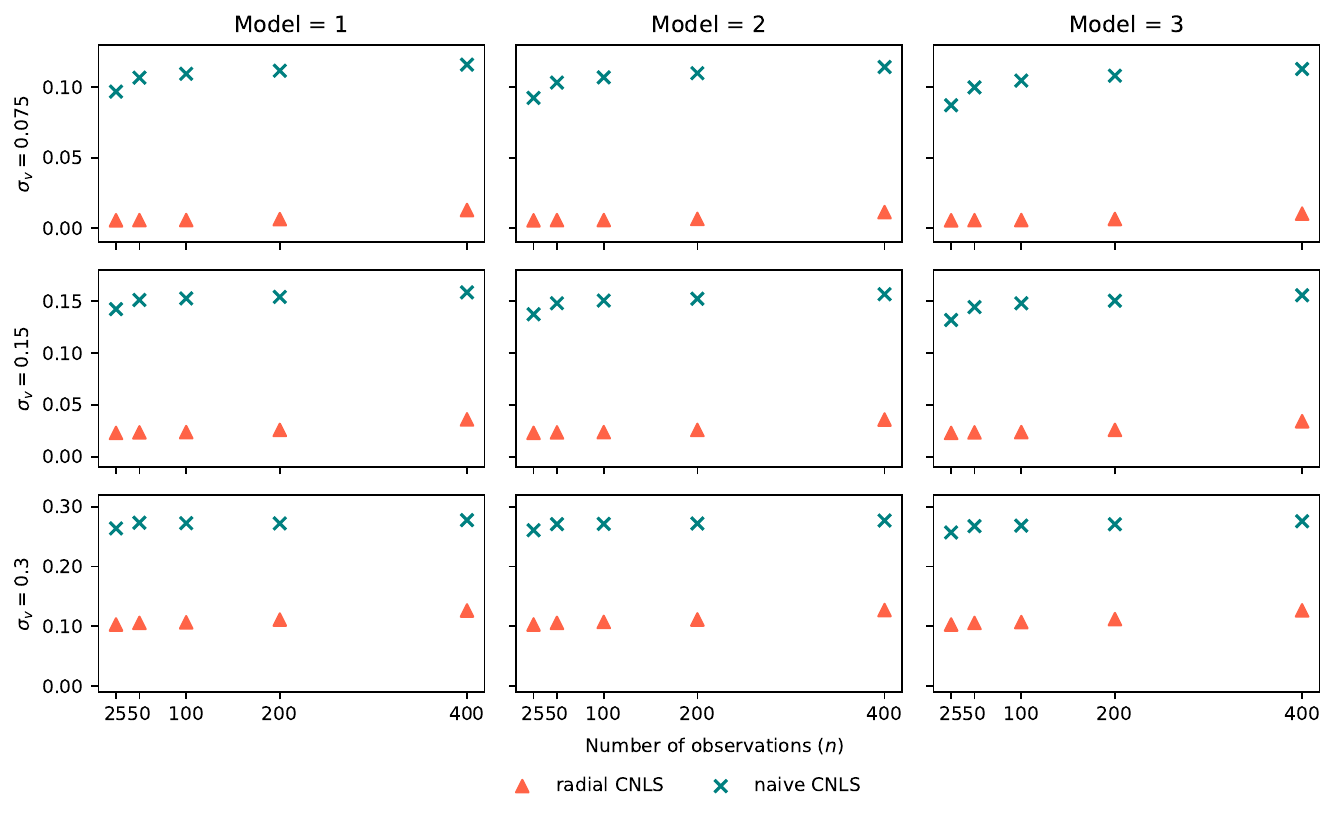}
    \caption{MSE results of radial CNLS and naive CNLS in DGP I with Type-A.}
    \label{fig:fig1}
\end{figure}

\begin{figure}[H]
    \centering
    \includegraphics[width=\textwidth]{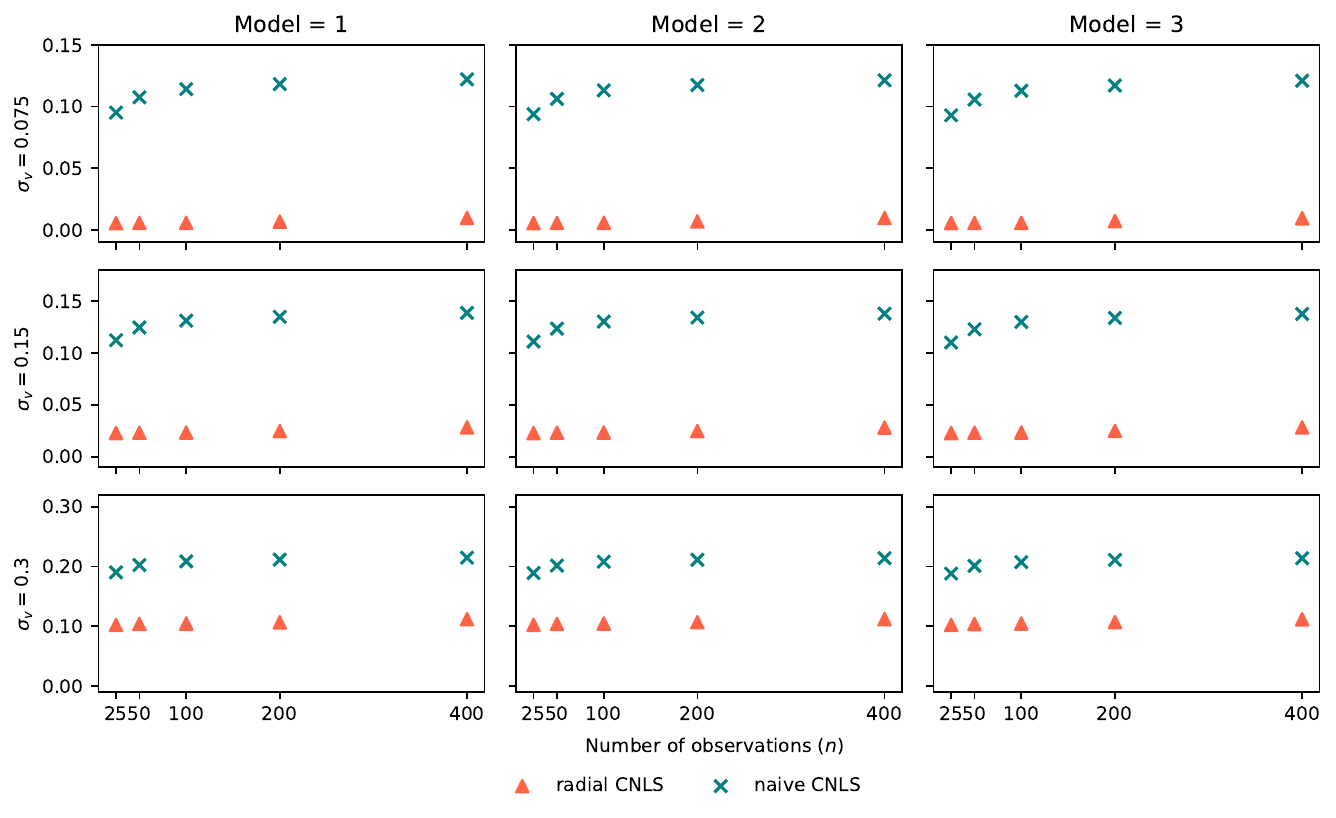}
    \caption{MSE results of radial CNLS and naive CNLS in DGP II.}
    \label{fig:fig2}
\end{figure}

Figs.~\ref{fig:fig1} and~\ref{fig:fig2} depict the MSE results of radial CNLS and naive CNLS in DGP I with Type-A and DGP II. Radial CNLS outperforms naive CNLS in all scenarios considered, suggesting that radial CNLS is a more efficient method for estimating the input distance function. In contrast to arbitrarily choosing the input, normalizing the input distance function can satisfy the orthogonality conditions, which probably helps fit the true function with smaller expected errors. Several other interesting findings are also observed.

First, the curvature of the generated production function is likely to affect the performance of estimators. For instance, compared to the other two models in DGPs I and II, Model 3 has the smallest MSE values due to that the generated production function is more concave in terms of the outputs. Second, the MSE values increase as the noise becomes large. It is clearly evident from Figs.~\ref{fig:fig1} and~\ref{fig:fig2} that the estimation of the input distance function becomes worse when $\sigma_v$ increases. Third, compared to DGP I, DGP II has a smaller MSE difference between the two estimators. This is because the generated dataset by DGP II is more balanced (see Figure 1 in \citeauthor{Fare2010}, \citeyear{Fare2010}), which dampens the disadvantage of the naive CNLS approach. Further, Figs.B1 and B2 in Appendix B demonstrate the MAD results of radial CNLS and naive CNLS in two sets of DGPs, showing similar findings as in the MSE comparison and supporting the main conclusion of experiment 1. 

Given the potential influence of the curvature of the true output isoquant, we further compare two CNLS approaches based on DGP I with Type-B, where a more balanced dataset in terms of outputs is generated. As shown in Table \ref{tab:type}, the MSE values of naive CNLS are always larger than those of radial CNLS.
\begin{table}[H]
  \centering
  \caption{DGP I with Type-B ($n=400$).}
  \footnotesize{
    \begin{tabular}{llrrrrr}
    \toprule
    \multicolumn{1}{c}{\multirow{2}[4]{*}{Model}} & \multicolumn{1}{c}{\multirow{2}[4]{*}{$\sigma_v$}} & \multicolumn{2}{c}{MSE} &       & \multicolumn{2}{c}{MAD} \\
\cmidrule{3-4}\cmidrule{6-7}          &       & \multicolumn{1}{l}{radial CNLS} & \multicolumn{1}{l}{naive CNLS} &       & \multicolumn{1}{l}{radial CNLS} & \multicolumn{1}{l}{naive CNLS} \\
    \midrule
    1     & 0.075 & 0.0122 & 0.1147 &       & 0.0725 & 0.2772 \\
          & 0.15  & 0.0358 & 0.1570 &       & 0.1375 & 0.3227 \\
          & 0.3   & 0.1266 & 0.2770 &       & 0.2662 & 0.4227 \\
    2     & 0.075 & 0.0101 & 0.1124 &       & 0.0684 & 0.2730 \\
          & 0.15  & 0.0331 & 0.1547 &       & 0.1337 & 0.3192 \\
          & 0.3   & 0.1242 & 0.2749 &       & 0.2634 & 0.4206 \\
    3     & 0.075 & 0.0088 & 0.1104 &       & 0.0656 & 0.2692 \\
          & 0.15  & 0.0310 & 0.1530 &       & 0.1309 & 0.3165 \\
          & 0.3   & 0.1222 & 0.2730 &       & 0.2618 & 0.4187 \\
    \bottomrule
    \end{tabular}%
    }
  \label{tab:type}%
\end{table}%

\subsection{Experiment 2}

We proceed to compare the performance among radial CNLS, DEA, and SFA. Following \citet{Schaefer2018}, we consider 180 scenarios in experiment 2 with different numbers of observations, $n \in \{25, 50, 100, 200, 400\}$, the noise-to-signal ratios, $\rho_{nts} \in \{0, 0.5, 1, 2\}$, where $\rho_{nts}=\sigma_v/\sigma_u$, and $\sigma_u=0.15$. Each scenario is also run 500 times to compute the MSE and MAD metrics. Note that the input-oriented DEA model with variable returns to scale and SFA with Cobb Douglas function (SFA-CD) and translog function (SFA-TL) specifications are considered in the benchmarking procedures. 

Tables \ref{tab:tab2} and B1 in Appendix B demonstrate the performance comparison of radial CNLS, DEA, and SFA with $n=400$. This experiment further confirms the superiority of the proposed radial CNLS approach in estimating input distance function compared to other available methods. Specifically, the radial CNLS approach has the lowest MSE and MAD values in all scenarios. Notably, in the case of the DGPs with inefficiency only, radial CNLS also works best compared to its counterparts. The MSE and MAD values of the SFA-CD approach are larger than those of others and increase much more sharply at the high noise-to-signal ratio of $\rho_{nts}=2$ (cf. \citeauthor{Schaefer2018}, \citeyear{Schaefer2018}). This is because when estimating the Cobb-Douglas function using SFA, the considered DGPs are misspecified, especially in polynomial technologies. Note that the results show that radial CNLS and DEA are relatively robust to polynomial and translog technologies. Furthermore, as expected, the values of MSE and MAD increase as the noise increases (see, e.g., \citeauthor{Henningsen2015}, \citeyear{Henningsen2015}; \citeauthor{Schaefer2018}, \citeyear{Schaefer2018}; \citeauthor{Ahn2023}, \citeyear{Ahn2023}). 
\vspace{-0.3cm}
\begin{table}[H]
  \centering
  \caption{Performance of radial CNLS, DEA, and SFA with $n=400$.}
  \footnotesize{
    \begin{tabular}{lllrrrrrrrrr}
    \toprule
    \multirow{2}[4]{*}{DGP} & \multirow{2}[4]{*}{Model} & \multirow{2}[4]{*}{$\rho$} & \multicolumn{4}{c}{MSE}       &       & \multicolumn{4}{c}{MAD} \\
\cmidrule{4-7}\cmidrule{9-12}          &       &       & \multicolumn{1}{l}{CNLS} & \multicolumn{1}{l}{DEA} & \multicolumn{1}{l}{SFA-TL} & \multicolumn{1}{l}{SFA-CD} &       & \multicolumn{1}{l}{CNLS} & \multicolumn{1}{l}{DEA} & \multicolumn{1}{l}{SFA-TL} & \multicolumn{1}{l}{SFA-CD} \\
    \midrule
    I     & 1     & 0     & 0.013 & 0.083 & 1.189 & 44.739 &       & 0.082 & 0.218 & 0.479 & 1.531 \\
    (Type-A)      &       & 0.5   & 0.017 & 0.140 & 2.294 & 48.761 &       & 0.100 & 0.317 & 0.589 & 1.552 \\
          &       & 1     & 0.032 & 0.222 & 2.443 & 16.478 &       & 0.140 & 0.430 & 0.709 & 1.364 \\
          &       & 2     & 0.099 & 0.380 & 8.255 & 64.802 &       & 0.240 & 0.574 & 1.145 & 2.003 \\
          & 2     & 0     & 0.012 & 0.075 & 6.225 & 60.838 &       & 0.082 & 0.205 & 0.642 & 1.923 \\
          &       & 0.5   & 0.017 & 0.127 & 6.404 & 60.657 &       & 0.099 & 0.298 & 0.747 & 2.000 \\
          &       & 1     & 0.033 & 0.209 & 3.853 & 12.270 &       & 0.142 & 0.414 & 0.846 & 1.332 \\
          &       & 2     & 0.099 & 0.369 & 3.986 & 60.876 &       & 0.241 & 0.566 & 1.025 & 1.883 \\
          & 3     & 0     & 0.012 & 0.069 & 1.036 & 59.532 &       & 0.081 & 0.194 & 0.536 & 1.935 \\
          &       & 0.5   & 0.017 & 0.117 & 1.656 & 61.946 &       & 0.099 & 0.283 & 0.671 & 2.089 \\
          &       & 1     & 0.032 & 0.197 & 2.092 & 112.842&       & 0.140 & 0.400 & 0.790 & 2.263 \\
          &       & 2     & 0.099 & 0.353 & 2.159 & 65.421 &       & 0.240 & 0.553 & 0.812 & 1.663 \\
    II    & 1     & 0     & 0.013 & 0.077 & 0.336 & 6.708  &       & 0.083 & 0.235 & 0.339 & 0.852 \\
          &       & 0.5   & 0.017 & 0.083 & 0.463 & 6.344  &       & 0.100 & 0.246 & 0.374 & 0.827 \\
          &       & 1     & 0.032 & 0.101 & 0.434 & 2.897  &       & 0.140 & 0.272 & 0.378 & 0.757 \\
          &       & 2     & 0.095 & 0.175 & 0.819 & 5.775  &       & 0.237 & 0.348 & 0.518 & 0.929 \\
          & 2     & 0     & 0.013 & 0.072 & 0.339 & 18.427 &       & 0.082 & 0.226 & 0.329 & 0.952 \\
          &       & 0.5   & 0.017 & 0.078 & 0.403 & 4.658  &       & 0.100 & 0.236 & 0.346 & 0.781 \\
          &       & 1     & 0.032 & 0.095 & 0.429 & 6.117  &       & 0.140 & 0.262 & 0.371 & 0.807 \\
          &       & 2     & 0.094 & 0.168 & 0.690 & 6.425  &       & 0.237 & 0.340 & 0.495 & 0.925 \\
          & 3     & 0     & 0.013 & 0.069 & 0.378 & 6.503  &       & 0.083 & 0.221 & 0.345 & 0.880 \\
          &       & 0.5   & 0.018 & 0.075 & 0.348 & 5.897  &       & 0.102 & 0.231 & 0.334 & 0.832 \\
          &       & 1     & 0.032 & 0.092 & 0.365 & 5.813  &       & 0.140 & 0.258 & 0.352 & 0.822 \\
          &       & 2     & 0.095 & 0.164 & 0.755 & 10.009 &       & 0.238 & 0.336 & 0.503 & 0.989 \\
    \bottomrule
    \end{tabular}%
    }
    \label{tab:tab2}%
\end{table}%

%-----------------%
% 
%-----------------%
\vspace{-0.5cm}
\section{Application}\label{sec:app}

The most significant real-world application of CNLS has thus far been incentive regulation of electricity distribution firms (\citeauthor{Kuosmanen2012b}, \citeyear{Kuosmanen2012b}). This literature introduces the CNLS estimator that combines the virtues of DEA and SFA to the Finnish electricity distribution network regulation. But most existing applications of CNLS or other frontier estimation techniques focus on either variable cost or total cost and hence have demonstrable shortcomings (\citeauthor{Kuosmanen2020d}, \citeyear{Kuosmanen2020d}). To consider variable cost and fixed cost simultaneously in the benchmark regulation, \citet{Kuosmanen2017a} propose a new CNLS approach with DDF (i.e., DDF CNLS), and \citet{Kuosmanen2020d} and \citet{Kuosmanen2022} further develop a CNLS approach to incorporate the input requirement function (i.e., IRF CNLS). 

However, the DDF CNLS approach requires a prespecified direction vector, and the IRF CNLS approach needs to project in the direction of one input. The estimates by DDF CNLS or IRF CNLS are not invariant to the direction vector or the input choice. This motivates us to propose and apply the radial CNLS approach to energy regulation practice, where radial CNLS does not rely on any direction vector and is immune to the input selection.

\subsection{Model specification}

Following \citet{Kuosmanen2020d}, we consider the following the multiplicative cost frontier model with contextual variables
\begin{equation}
     \label{eq:cost}
     \bx_{i,t} = C(\by_{i,t}, \bb_{i,t})\exp(\blambda \bz_{i,t} + \varepsilon_{i,t})
\end{equation}
where $C$ is a non-decreasing convex conical cost function. $\bx$, $\by$, and $\bb$ are inputs, desirable outputs, and undesirable outputs, respectively. $\bz$ represents a set of contextual variables, and $\blambda$ denotes the associated marginal coefficients. $\varepsilon$ is a composite error term involving inefficiency and random effects. Accordingly, the input, output, and contextual variables of the cost frontier model~\eqref{eq:cost} are specified as follows.
\setlist{nolistsep}
\begin{itemize}[noitemsep]
    \item[] \textbf{Inputs} ($\bx$)
    \begin{itemize}
        \item[] $x_1 =$ Fixed cost, capital stock (regulatory asset value, NKA, \officialeuro)
        \item[] $x_2 =$ Variable cost, controllable operational expenditure (KOPEX, \officialeuro)
    \end{itemize}
    \item[] \textbf{Outputs} 
        \begin{itemize}
            \item[] Desirable outputs ($\by$) 
            \begin{itemize}
                \item[] $y_1 =$ Energy supply (GWh, weighted by voltage)
                \item[] $y_2 =$ Network length (km)
                \item[] $y_3 =$ Number of use points
            \end{itemize}
            \item[] Undesirable output ($b$) 
            \begin{itemize}
                \item[] $b_1 =$ Outages (hedonic damage cost, \officialeuro)
            \end{itemize}
        \end{itemize}
    \item[] \textbf{Contextual variables} ($\bz$)  
    \begin{itemize}
        \item[] $z_1 =$ Connection points / Use points
        \item[] $z_2 =$ Energy loss (\%)
    \end{itemize}
\end{itemize}

\vspace{0.3cm}
While radial CNLS~\eqref{eq:cnls_raidal} is ready for estimating the multiple input-multiple output cost frontier model~\eqref{eq:cost}, it might overfit training data and predictably perform poorly on testing data.\footnote{
    Overfitting is a longstanding problem in convex regression and other general nonparametric regression, where the subgradients may become very large at the boundary of the convex hull of the design points (\citeauthor{Liao2023}, \citeyear{Liao2023}).
}
Considering that the benchmark regulation generally involves future economic incentives, we set a lower bound and upper bound on each subgradient to control their magnitudes to increase the performance of radial CNLS (\citeauthor{Kuosmanen2022}, \citeyear{Kuosmanen2022}). Other restrictions on fitted subgradients for reducing overfitting include the Lipchitz norm (e.g., \citeauthor{Mazumder2019}, \citeyear{Mazumder2019}) and $L_2$ norm (e.g., \citeauthor{Dai2023c}, \citeyear{Dai2023c}). One could even respecify the loss function to mitigate the effect of overfitting (e.g., \citeauthor{Liao2023}, \citeyear{Liao2023}). 

After taking into account these specifications and choosing $x_1$ as numeraire, we rephrase radial CNLS~\eqref{eq:cnls_raidal} to estimate cost frontier model~\eqref{eq:cost} and have the following nonlinear programming problem
\begin{alignat}{2}
\underset{\chi, \bbeta, \bgamma, \blambda, \delta, \mu, \varepsilon} {\mathop{\min }}\, \quad & \sum\limits_{t=1}^{T}\sum\limits_{i=1}^{N}\varepsilon_{it}^2 &{\quad}& \label{eq:cnls_app}\\
\mbox{s.t.}\quad
& \ln x_{1it} = - \ln (\chi_{it}) + \delta(\ln x_{1it} - \ln x_{2it}) + \blambda^{\prime}\bz_{it} +\varepsilon_{i,t}  &{}& \forall i,t\notag \\
& \chi_{it} = \bbeta^{\prime}_{i,t}  (\bx_{it}/(x_{1it}^{1/2}x_{2it}^{1/2})) + \mu_{it} b_{1it} - \bgamma^{\prime}_{it} \by_{it} &{}& \forall i, t\notag \\
& \chi_{it} \le \bbeta^{\prime}_{hs} (\bx_{it}/(x_{1it}^{1/2}x_{2it}^{1/2})) + \mu_{hs} b_{1it} - \bgamma^{\prime}_{hs} \by_{it} &{}& \forall i,h; \forall t,s \notag  \\ 
& \bbeta_{lo} \le \bbeta_{it} \le \bbeta_{up} \,,\ \bgamma_{lo} \le \bgamma_{it} \le \bgamma_{up} \,,\ \mu_{lo} \le \mu_{it} \le \mu_{up}  &{}& \forall i,t \notag  \\  
& \bbeta_{it} \ge \textbf{0} \,,\ \bgamma_{it} \ge \textbf{0} &{}& \forall i,t\notag  
\end{alignat}
where the fourth set of constraints refers to the weight restrictions on subgradients, and $\mu_{it}$ denotes the marginal coefficient of the outages. The lower and upper bounds (e.g., $\bbeta_{lo}$ and $\bbeta_{up}$) in problem~\eqref{eq:cnls_app} can be determined by either a data-driven approach (e.g., cross-validation) or decision-makers and/or stakeholders. A major practical advantage of the latter is that the weight restrictions can be easily communicated to decision-makers (e.g., the regulator) and stakeholders (e.g., the regulated firms and their customers) as they can literally see the weight restrictions themselves and comment if those are too loose, too restrictive, or just fine. 

In practice, we first solve problem~\eqref{eq:cnls_app} without weight restrictions (i.e., the fourth sets of constraints) to obtain the subgradients estimates (i.e., $\hat{\bbeta}$, $\hat{\bgamma}$, and $\hat{\mu}$) and then take 10\% and 90\% quartiles of each subgradient as its lower and upper bounds to reestimate problem~\eqref{eq:cnls_app}.

\subsection{Empirical results}

In this section we discuss what degree of difference in the estimates with and without weight restrictions and compare radial CNLS with its alternative. We thus apply the proposed radial CNLS approach to a panel of 77 Finnish electricity distribution firms in the years 2008--2020.\footnote{
    The original dataset is applied to carry out the incentive regulation for Finnish electricity distribution networks (\citeauthor{Kuosmanen2022}, \citeyear{Kuosmanen2022}), and its earlier version has been widely used in, e.g., \citet{Kuosmanen2012b}, \citet{Kuosmanen2013}, and \citet{Kuosmanen2020d}.
}
The descriptive statistics for inputs, outputs, and contextual variables are summarized in Table B2 (Appendix B). See \citet{Kuosmanen2020d} for a detailed introduction to these selected variables. 

Table~\ref{tab:tab4} summarizes the results estimated by the basic radial CNLS model and weight-restricted radial CNLS model. Note that the estimated input distance function is a piece-wise linear function consisting of hyperplanes characterized by subgradients $\bbeta_i$, $\bgamma_i$, and $\mu_i$. When weight constraints are introduced to the radial CNLS model, the average estimates for the nonparametric part dramatically decrease, and the corresponding standard deviations also decline. This implies that a small part of firms has extremely estimated shadow prices (i.e., $\hat{\bbeta}_i$, $\hat{\bgamma}_i$, or $\hat{\mu}_i$), which can severely affect the fitted input distance function in a testing set and hence deteriorate the performance of economic incentives in energy regulation.

\begin{figure}[H]
    \centering
    \includegraphics[width=0.8\linewidth]{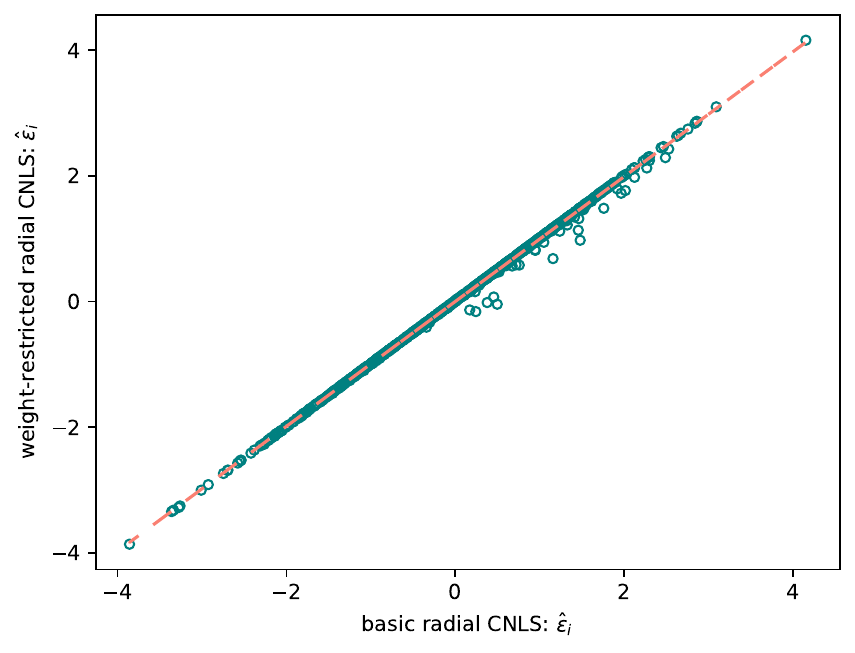}
    \caption{Estimated residuals: radial CNLS and weighted-restricted radial CNLS.}
    \label{fig:res}
\end{figure}

By comparing the 10\%, 50\%, and 90\% quartile estimates, we observe that the changes in estimated shadow prices remain rather marginal, suggesting that most of the firms would not be highly affected by additional weight constraints. Furthermore, the estimated residual comparison also shows little impact of weight restrictions on the firm's relative performance in that most firms cluster around the 45-degree line (see Fig.~\ref{fig:res}). That is, while the applied weight-restricted CNLS model will affect the estimated input distance function, it can avoid extreme shadow prices and show its robustness in inefficiency estimation and endogeneity bias.

It is worth noting that the minimal estimated $\hat{\mu}_i$ is negative in the basic radial CNLS model but is positive in the weight-restricted radial CNLS model. Unsurprisingly, the marginal effect of outages at the firm level can be positive or negative in the context of the benchmark regulation (\citeauthor{Kuosmanen2020d}, \citeyear{Kuosmanen2020d}). Recall that there is huge heterogeneity between Finnish electricity distribution firms in terms of variable and fixed costs as reflected by their standard deviation (see Table B2 in Appendix B). Firms with low outages can use a high level of quality and reliability of distribution as a competitive advantage in the yardstick competition. For these firms, the shadow price of outages is negative. Firms that face exceptionally severe weather shocks have higher operational costs due to large outages, deducing the positive shadow price of outages. This is because the standard compensations paid to customers for interruptions are also included in the variable cost. In practice, the threshold value for a positive outage effect is very high because the outage values must be exceptionally high compared to other firms such that the firm's operations appear competitive in the benchmark regulation between firms. Therefore, the impact of outages on the distribution firms is a ``U-shaped'' curve.
\begin{table}[H]
  \centering
  \caption{Radial CNLS estimates for the Finnish electricity distribution firms.}
  \footnotesize{
    \begin{tabular}{lrrrrrrr}
    \toprule
    & \multicolumn{1}{c}{Mean} & \multicolumn{1}{c}{Std. Dev.} & \multicolumn{1}{c}{Min.} & \multicolumn{1}{c}{p10} & \multicolumn{1}{c}{p50} & \multicolumn{1}{c}{p90} & \multicolumn{1}{c}{Max.} \\
    \midrule
    \multicolumn{8}{c}{Basic radial CNLS model: Problem~\eqref{eq:cnls_app} without weight constraints}  \\
    \textit{Nonparametric part} & & & & & & & \\
    $\hat{\beta}_1$                                         & 95.293   & 749.698   & 0.000   &  0.599  & 0.601   &  0.605   & 11511.607 \\
    $\hat{\beta}_2$                                         & 329.330  & 2413.736  & 0.000   &  0.000  & 0.000   &  0.088   & 22924.910 \\
    $\hat{\gamma}_1$(\officialeuro {\hskip0.05em} cents/kWh)& 28046.330& 294777.200& 0.000   &  0.001  & 0.186   &  382.347 & 5635193.000 \\
    $\hat{\gamma}_2$(\officialeuro/km)                      & 3937.605 & 41107.930 & 0.000   &  0.001  & 22.846  & 49.378   & 752614.085 \\
    $\hat{\gamma}_3$(\officialeuro/customer)                & 553.660  & 6175.607  & 0.000   &  0.000  & 0.672   & 3.978    & 130694.500  \\
    $\hat{\mu}$                                             & 49027.060& 675278.700& -139973.600 & 2.840  & 53.010  &224.093 & 14844690.000 \\
    \textit{Parametric part} & & & & & & &\\
    $\hat{\delta}$           & \multicolumn{7}{c}{4.10}  \\
    $\hat{\lambda}_1$        & \multicolumn{7}{c}{-0.71} \\
    $\hat{\lambda}_2$        & \multicolumn{7}{c}{4.11}  \\
    &  \\
    \multicolumn{8}{c}{Weight-restricted radial CNLS model: Problem~\eqref{eq:cnls_app}}  \\
    \textit{Nonparametric part} & & & & & & & \\
    $\hat{\beta}_1$                                         & 0.601    & 0.002    & 0.599   &  0.599  & 0.601   &  0.603   & 0.605 \\
    $\hat{\beta}_2$                                         & 0.011    & 0.025    & 0.000   &  0.000  & 0.000   &  0.050   & 0.088 \\
    $\hat{\gamma}_1$(\officialeuro {\hskip0.05em} cents/kWh)& 80.904   & 134.517  & 0.001   &  0.007  & 2.057   &  364.635 & 382.347 \\
    $\hat{\gamma}_2$(\officialeuro/km)                      & 21.685   & 16.134   & 0.001   &  0.010  & 22.817  & 46.901   & 49.378 \\
    $\hat{\gamma}_3$(\officialeuro/customer)                & 1.540    & 1.654    & 0.000   & 0.000   & 0.543   & 3.816    & 3.978  \\
    $\hat{\mu}$                                             & 64.726   & 61.222   & 2.840   & 2.881   & 64.738  & 141.897  & 224.093 \\
    \textit{Parametric part} & & & & & & &\\
    $\hat{\delta}$           & \multicolumn{7}{c}{4.10}  \\
    $\hat{\lambda}_1$        & \multicolumn{7}{c}{-0.71} \\
    $\hat{\lambda}_2$        & \multicolumn{7}{c}{3.98}  \\
    \bottomrule
    \end{tabular}%
    }
  \label{tab:tab4}%
\end{table}%

Furthermore, additional weight constraints also influence the estimated coefficient of the second contextual variable, energy loss. In the weight-restricted radial CNLS model, the value of the estimated coefficient of energy loss is 3.98, slightly lower than the value of 4.11 obtained from the basic radial CNLS model. Since the share of energy loss is on average higher in sparsely populated areas than in densely populated areas, the distorting effect of possible efficiency differences can be partially canceled out from the energy loss estimate with the help of weight restrictions. 

\begin{figure}[H]
	\centering
	\begin{subfigure}[b]{0.495\textwidth}
		\centering
		\includegraphics[width=1\textwidth]{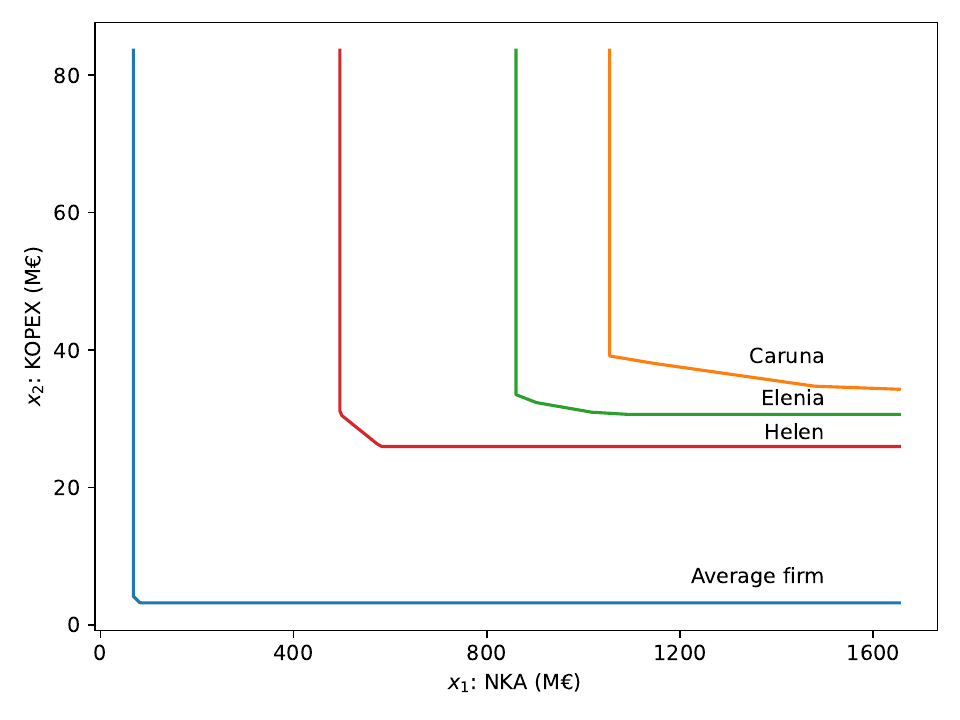} 
		\caption[]%
		{{\small Radial CNLS}}    
		\label{fig5:a}
	\end{subfigure}
	%\hfill
	\begin{subfigure}[b]{0.495\textwidth}  
		\centering 
		\includegraphics[width=1\textwidth]{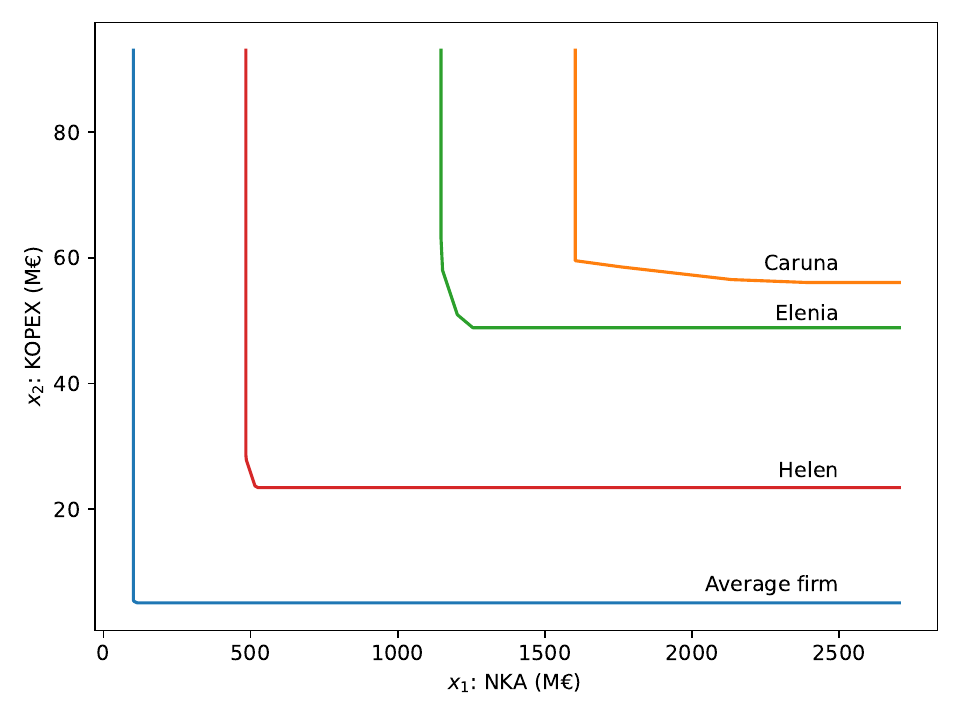} 
		\caption[]%
		{{\small IRF CNLS}}    
		\label{fig5:b}
	\end{subfigure}
	\caption[]%
	{\small Estimated input isoquants of the average firm and the three largest firms.} 
	\label{fig:isoquant}
\end{figure}

Fig.~\ref{fig:isoquant} demonstrates the input isoquants of the three largest distribution firms (Caruna, Elenia, and Helen) and the average level of all firms based on radial CNLS and IRF CNLS estimates. The horizontal axis is fixed cost (NKA), and the vertical axis is variable cost (KOPEX): both are rescaled in Million \officialeuro. Note that the shape of the input isoquant is determined by the output structure of the firm in the multiple output setting. Given the output structure of the three largest distribution firms, both two subfigures illustrate the considerable substitution possibilities between NKA and KOPEX. But the right figure shows that the average firm yields almost a Leontief-type input isoquant given the averaged output structure. Furthermore, it can be seen from Fig.~\ref{fig:isoquant} that the input isoquants become more curved (i.e., more substitution possibilities) when we use two inputs in the radial CNLS model rather than projecting in the direction of one input in the IRF CNLS model.

%-----------------%
% 
%-----------------%

\section{Conclusions}\label{sec:concl}

Modeling of joint production is a more vexing problem than most authors realize. The axiomatic nonparametric approach is the only theoretically sound alternative when data are perturbed by random noise, and the parametric approach is likely to violate the theoretical properties of the distance function. Modeling radial input and output distance functions in CNLS/StoNED has also been an enigma for almost two decades. Finally, in this paper the problem has been solved in the case of the input distance function, but the output distance function partially remains.

This paper proposes a new radial CNLS approach to estimate the input distance function in the multiple input-multiple output specification. We demonstrate how to transform the input distance function correctly and show that the orthogonality conditions can be satisfied in the developed approach. We then discuss the possible extensions from the other three aspects: considering the general case with more than two inputs, connecting with the efficiency analysis, and estimating the output distance function. 

Our simulations compare the finite sample performance of radial CNLS and other deterministic and stochastic frontier approaches regarding the input distance function estimation. Radial CNLS outperforms the naive CNLS, DEA, and SFA approaches in all scenarios considered, suggesting that radial CNLS is a more efficient method for estimating the input distance function.

The developed approach is further applied to the Finnish electricity distribution network regulation. To reduce the potential overfitting and increase the out-of-sample performance in energy regulation, we introduce weight restrictions on shadow prices to the framework of radial CNLS. In contrast to other traditional model specifications (e.g., DDF and IRF), radial CNLS does not rely on any direction vector and is immune to the input selection. The illustrated input isoquants suggest more substitution possibilities.

Future research could further examine joint production from the perspectives of nonparametric identification (see, e.g., \citeauthor{Matzkin2007}, \citeyear{Matzkin2007}; \citeauthor{Chiappori2015}, \citeyear{Chiappori2015}; \citeauthor{Heckman2010}, \citeyear{Heckman2010}) and partial identification (see, e.g., \citeauthor{Manski2009}, \citeyear{Manski2009}; \citeauthor{Tamer2010}, \citeyear{Tamer2010}). Axioms of production theory (linear homogeneity) provide moment conditions (i.e., generalized method of moments), but the model remains inherently underidentified. Underidentification is conventionally seen as a failure in econometrics (\citeauthor{Arellano2012}, \citeyear{Arellano2012}), but in many relevant applications, partial identification may be sufficient for the task at hand. One possible avenue of future research is to impose moment conditions supported by theory (linear homogeneity) and apply additional statistical criteria (e.g., maximum likelihood and least squares) to achieve (partial) identification. 

%-----------------%
% 
%-----------------%

\section*{Acknowledgments}\label{sec:ack}

The authors wish to acknowledge CSC – IT Center for Science, Finland, for computational resources. Sheng Dai gratefully acknowledges financial support from the OP Group Research Foundation [grant no.~20230008] and the Turku University Foundation [grant no.~081520].

%-----------------%
% 
%-----------------%

\baselineskip 12pt
\bibliographystyle{dcu}
\bibliography{References.bib}

%-----------------%
% 
%-----------------%

\clearpage
\newpage
\baselineskip 20pt
\section*{Online Supplement}

%-----------------%
% 
%-----------------%

\renewcommand{\thesubsection}{Appendix \Alph{subsection}:}
\renewcommand{\theequation} {A.\arabic{equation}}
\setcounter{equation}{0}
\renewcommand{\thetable}{B\arabic{table}}
\setcounter{table}{0}
\renewcommand{\thefigure}{B\arabic{figure}}
\setcounter{figure}{0}
\renewcommand{\thetheorem}{\arabic{theorem}}
\setcounter{theorem}{0}

\subsection{Proofs of Theorems}\label{sec:proof}

\begin{theorem}
If the observed data are generated according to the data generating process (DGP) described, then the value of the input distance function is equal to
	\begin{equation*}
		D_i(\bx_i, \by_i) = \exp(-\varepsilon_i)
	\end{equation*}
\end{theorem}
\begin{proof}
If the observed data are generated according to the DGP stated in Section 2.1, then we have 
\begin{equation*}
    D_i(\bx, \by) = D_i(\exp(\varepsilon_i)\bx^*, \by)
\end{equation*}

By utilizing the linear homogeneity of input distance function $D_i$, the above input distance function $D_i(\exp(\varepsilon_i)\bx^*, \by)$ can be reorganized as 
\begin{equation*}
     D_i(\bx, \by) = D_i(\exp(\varepsilon_i)\bx^*, \by) = \exp(\varepsilon_i)^{-1} D_i(\bx^*, \by)
\end{equation*}
By definition, the efficient input-output vectors must be on the boundary of the production possibility set $T$ and satisfy the condition
$D_i(\bx^*, \by) \equiv 1$. We thus have 
\begin{equation*}
     D_i(\bx, \by) = \exp(\varepsilon_i)^{-1}
\end{equation*}
\end{proof} 

\begin{theorem}
The radial CNLS residuals obtained as the optimal solution to problem (10) satisfy the following sample orthogonality condition
	\begin{equation*}
		\sum_{i=1}^{n}(\ln x_{1i} - \ln x_{2i})\cdot \hat{\varepsilon}_i=0
	\end{equation*}
\end{theorem}
\begin{proof}
The Lagrangian function of problem (10) can be formulated
\begin{alignat*}{2}
L(\blambda,\balpha,\bbeta,\bmu,{\boldsymbol \eta}, {\boldsymbol \rho})&=\sum\limits_{i=1}^{n}\varepsilon _i^2+\sum\limits_{i=1}^{n}\mu_i(\ln x_{1i}+\ln (\chi_i) - \delta(\ln x_{1i} - \ln x_{2i}) - \varepsilon_{i})  \\
&+\sum\limits_{i=1}^{n}{\sum\limits_{h=1}^{n}{{\lambda_{ih}}(\alpha_i + \bbeta^{\prime}_i (\bx_i/(x_{1i}^{1-d}x_{2i}^d)) - \bgamma^{\prime}_i \by_i - \alpha_h - \bbeta^{\prime}_{h}  (\bx_i/(x_{1i}^{1-d}x_{2i}^d)) + \bgamma^{\prime}_{h}\by_i)}} \\
&-\sum\limits_{i=1}^{n}{\boldsymbol \eta}{'}_i\bbeta_i -\sum\limits_{i=1}^{n}{\boldsymbol \rho}{'}_i\bgamma_i
\end{alignat*}
where $\mu_i$, $\lambda_{ih}$, $\eta_i$, and $\rho$ are the Lagrangian multipliers of each constraints. The optimal solution of problem (8) satisfies the first order conditions with respect to $\varepsilon_i$ and $\delta$
\begin{align}
& \frac{\partial L}{\partial{\varepsilon_i}}=2\hat{\varepsilon}_i-\mu_i=0\text{  } \forall i \notag \\
& \frac{\partial L}{\partial{\delta}}=\sum\limits_{i=1}^{n}\mu_i(\ln x_{1i} - \ln x_{2i})=0 \notag
\end{align}
Combining these two first order conditions, we thus have
	\begin{equation*}
		\sum_{i=1}^{n}(\ln x_{1i} - \ln x_{2i})\cdot \hat{\varepsilon}_i=0
	\end{equation*}
\end{proof} 

\newpage
\subsection{Additional tables and figures}

\begin{figure}[H]
    \centering
    \includegraphics[width=0.92\textwidth]{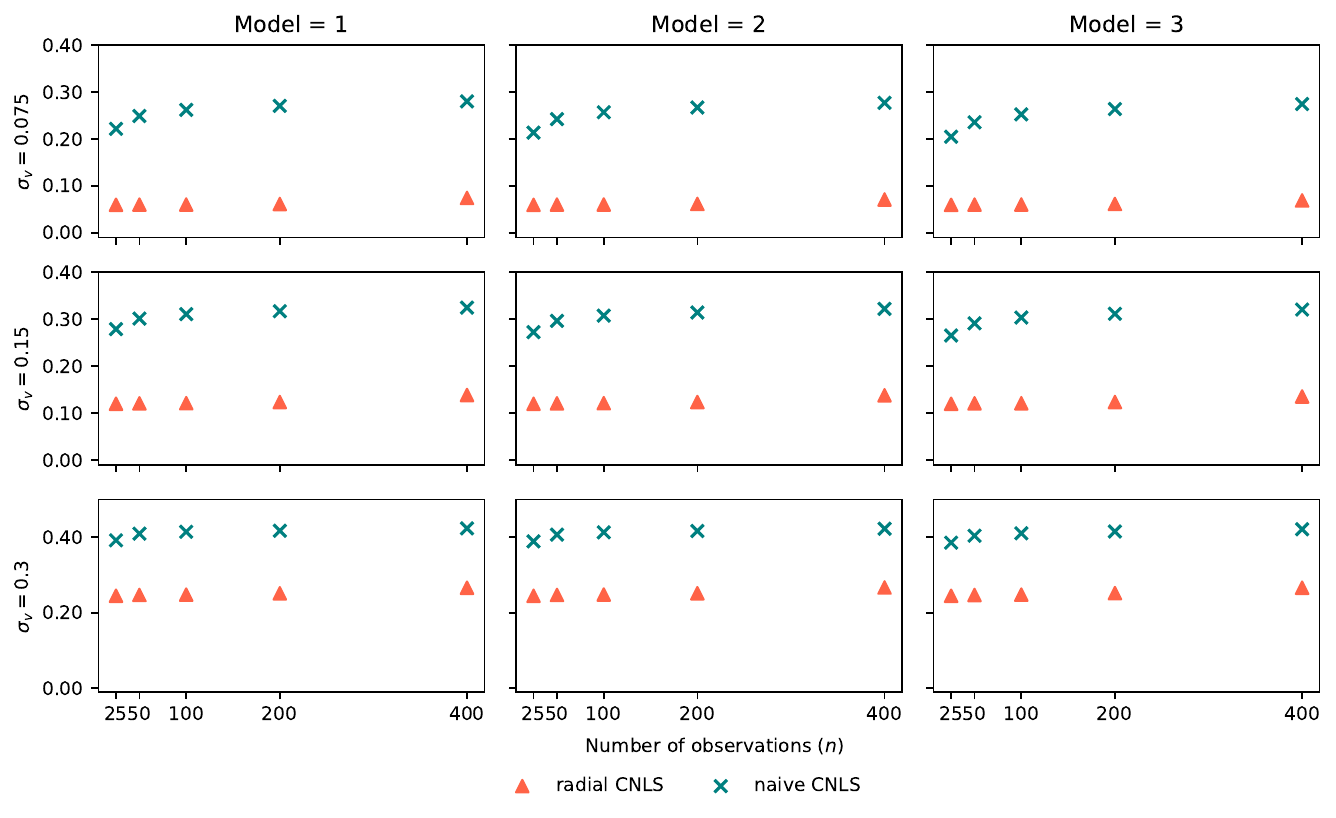}
    \caption{Mean absolute errors of radial CNLS and naive CNLS in DGP I with Type-A.}
    \label{fig:a1}
\end{figure}
\vspace{-0.5cm}

\begin{figure}[H]
    \centering
    \includegraphics[width=0.92\textwidth]{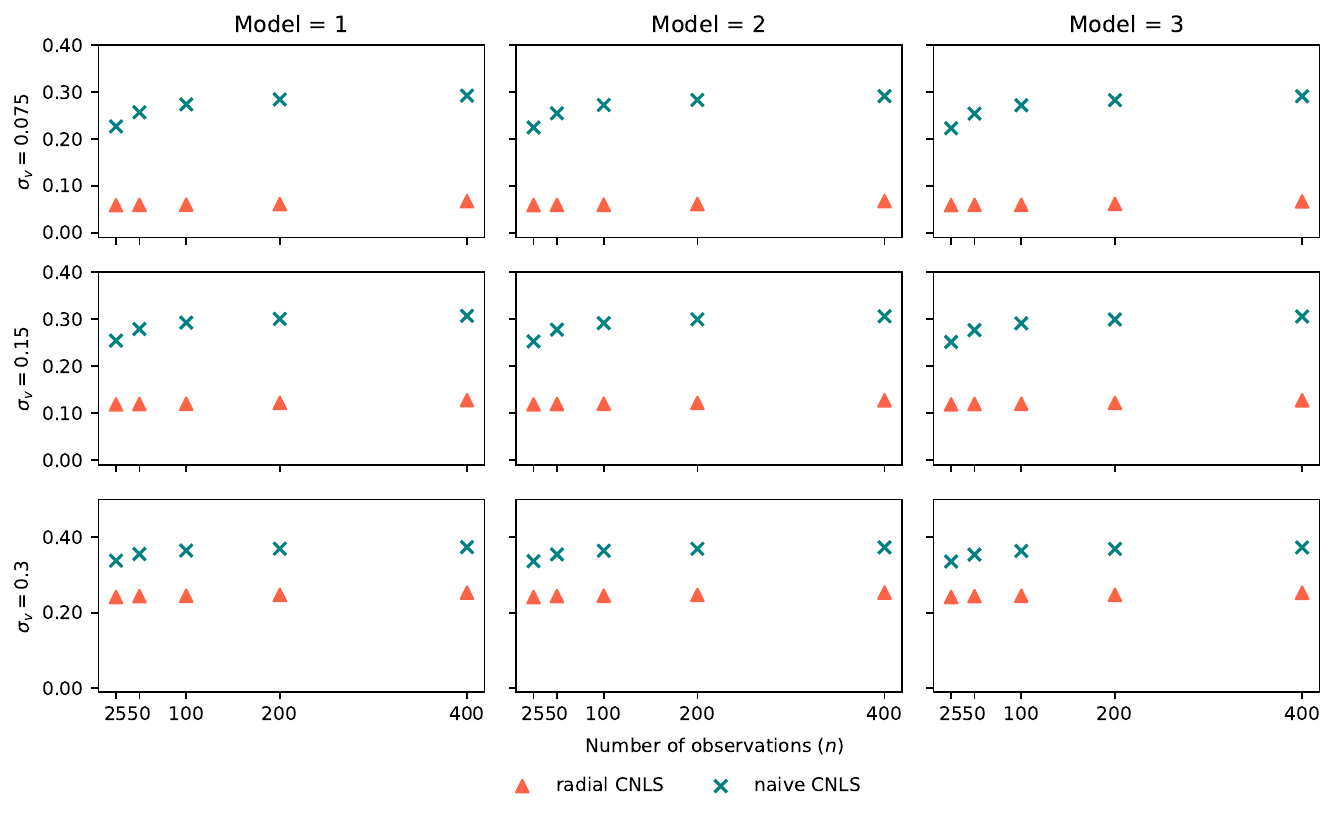}
    \caption{Mean absolute errors of radial CNLS and naive CNLS in DGP II.}
    \label{fig:a2}
\end{figure}

\begin{table}[H]
  \centering
  \caption{Performance of radial CNLS, DEA, and SFA with $n=400$ in DGP I with Type-B.}
    \small{
    \begin{tabular}{crrrrrrrrrr}
    \toprule
    \multicolumn{1}{c}{\multirow{2}[4]{*}{Model}} & \multicolumn{1}{c}{\multirow{2}[4]{*}{$\rho$}} & \multicolumn{4}{c}{Mean Squared Error}       &       & \multicolumn{4}{c}{Mean Absolute Error} \\
\cmidrule{3-6}\cmidrule{8-11}          &       & \multicolumn{1}{l}{CNLS} & \multicolumn{1}{l}{DEA} & \multicolumn{1}{l}{SFA-TL} & \multicolumn{1}{l}{SFA-CD} &       & \multicolumn{1}{l}{CNLS} & \multicolumn{1}{l}{DEA} & \multicolumn{1}{l}{SFA-TL} & \multicolumn{1}{l}{SFA-CD} \\
    \midrule
    1     & 0     & 0.013 & 0.076 & 0.508 & 12.350 &       & 0.082 & 0.234 & 0.189 & 0.859 \\
           & 0.5   & 0.017 & 0.081 & 0.499 & 11.941 &       & 0.100 & 0.242 & 0.195 & 0.862 \\
           & 1     & 0.030 & 0.097 & 0.500 & 12.779 &       & 0.138 & 0.266 & 0.210 & 0.862 \\
           & 2     & 0.094 & 0.169 & 0.797 & 15.031 &       & 0.237 & 0.341 & 0.301 & 0.895 \\
    2     & 0     & 0.012 & 0.070 & 0.531 & 13.279 &       & 0.082 & 0.223 & 0.200 & 0.933 \\
           & 0.5   & 0.017 & 0.076 & 0.668 & 13.204 &       & 0.101 & 0.233 & 0.213 & 0.924 \\
           & 1     & 0.030 & 0.092 & 0.812 & 15.970 &       & 0.138 & 0.257 & 0.242 & 0.933 \\
           & 2     & 0.094 & 0.164 & 1.742 & 16.856 &       & 0.238 & 0.335 & 0.388 & 0.969 \\
    3     & 0     & 0.012 & 0.068 & 1.328 & 15.575 &       & 0.082 & 0.218 & 0.263 & 1.028 \\
           & 0.5   & 0.017 & 0.073 & 1.472 & 15.852 &       & 0.100 & 0.228 & 0.282 & 1.015 \\
           & 1     & 0.030 & 0.089 & 1.701 & 15.810 &       & 0.138 & 0.253 & 0.329 & 1.012 \\
           & 2     & 0.094 & 0.160 & 2.734 & 17.903 &       & 0.237 & 0.331 & 0.500 & 1.041 \\
    \bottomrule
    \end{tabular}%
    }
  \label{tab:a2}%
\end{table}%

\begin{table}[H]
  \centering
  \caption{The descriptive statistics for inputs, outputs, and contextual variables.}
     \small{
    \begin{tabular}{lrrrrr}
    \toprule
    Variable & \multicolumn{1}{l}{Mean} & \multicolumn{1}{l}{Std. Dev.} & \multicolumn{1}{l}{Min.} & \multicolumn{1}{l}{Max.} & \multicolumn{1}{l}{Obs.} \\
    \midrule
Capital stock ($x_1$)   & 126253.5 & 284279.9 & 1540.6 & 2664448.0 & 1001 \\
Controllable operational expenditure ($x_2$)   & 5864.9 & 11156.8 & 115.2 & 114993.4 & 1001 \\
Energy supply ($y_1$)  & 569.3 & 1095.1 & 15.8  & 7528.9 & 1001 \\
Network length ($y_2$)    & 5063.1 & 11832.3 & 133.6 & 80278.2 & 1001 \\
Number of use points ($y_3$)    & 44721.0 & 83271.3 & 741.0 & 479365.0 & 1001 \\
Outages ($b_1$)    & 2202.7 & 8520.3 & 2.3   & 173568.5 & 1001 \\
Connection points / Use points ($z_1$)   & 0.63  & 0.24  & 0.09  & 1.00  & 1001 \\
Energy loss ($z_1$)   & 0.04  & 0.03  & 0.00  & 0.24  & 1001 \\
    \bottomrule
    \end{tabular}%
    }
  \label{tab:des}%
\end{table}%

\end{document}